# The origin of the distinction between microscopic formulas for stress and Cauchy stress

Youping Chen
Department of Mechanical and Aerospace Engineering, University of Florida, Gainesville, 32611

**Abstract** - Stress is calculated routinely in atomistic simulations. The widely used microscopic stress formulas derived from classical or quantum mechanics, however, are distinct from the concept of Cauchy stress, i.e., the true mechanical tress. This work examines various atomistic stress formulations and their inconsistencies. Using standard mathematic theorems and the law of mechanics, we show that Cauchy stress results unambiguously from the definition of internal force density, thereby removing the long-standing confusion about the atomistic basis of the fundamental property of Cauchy stress, and leading to a new atomistic formula for stress that has clear physical meaning and well-defined values, satisfies conservation law, and is fully consistent with the concept of Cauchy stress.

## I. INTRDUCTION

Energy, force, stress are basic concepts in the characterization of the state of condensed matter. Stress, in particular, is a key concept that links theory, simulation, and experiment. It has also been a subject of theoretical interest, as it can be used to establish correspondence between classical and quantum mechanics and between particle and continuum mechanics. Both the classical and quantum mechanical virial theorems show that the *system-wide* average stress in many-body systems is determined by the kinetic energy and the virial of the potential,[1-8] generally referred to as the kinetic and potential part of stress, respectively. In contrast, the atomistic formula for *local* stress remains a subject of debate. The critical issue is that no correspondence has been established between atomistic formulas for local stress and the fundamental concept of Cauchy stress.[8-11]

Cauchy stress, also known as the true mechanical stress, is defined as the force that the material on one side of a surface element exerts on the material on the other side, divided by the area of the surface.[12] It is a measure of the intensity of internal forces, has a clear physical origin, is the actual physical quantity measured in experiments, and is applicable on all scales. Atomistic stress formulas, on the other hand, were derived from classical or quantum mechanics as a function of the forces and positions of atoms, and is what have been used in *ab initio* calculations[13] as an intrinsic property of the quantum-mechanical ground state of matter, or in classical molecular dynamics or coarse-grained atomistic simulations to predict strength[14], fracture toughness[15, 16], hardness[17], or to quantify the effect of local stress on ferroelectricity[18], thermal conductivity[19, 20], phase transition[21, 22], etc.

There are numerous computational efforts that have attempted to understand the difference between various atomistic formulas for local stress and that between atomistic stress and Cauchy stress[8-11, 19, 23-25]. For example, FIG. 1 compares the stress at a dislocation core predicted by atomistic simulations using two most popular atomistic formulas with that by classical elasticity. While both atomistic stress formulas predict a zero stress at the dislocation core, the classical elasticity predicts an infinite stress. These computational efforts have enhanced our understanding of the significance of the distinction between atomistic formulas for stress and the Cauchy stress, the origin of the distinction, however, remains unidentified.

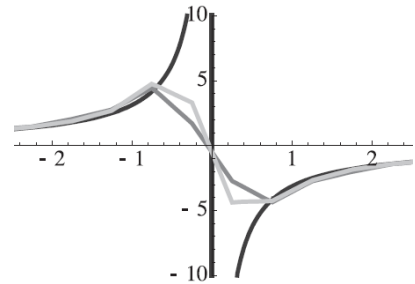

FIG. 1 Stress fields (vertical axis, in units of GPa) as a function of distance from an edge dislocation core in EAM Al (horizontal axis, in units of nm). Curves are plotted for molecular dynamics (MD) simulation results using virial stress (grey), Hardy's stress (light), and the mechanical stress calculated from elasticity theory (black).[25]

This work examines various formulations of microscopic stress and their inconsistencies. Using standard mathematic theorems, we identify the origin of this distinction and show that Cauchy stress results unambiguously from the definition of internal force density, thus leading to a new stress formula that unifies atomistic and Cauchy stress. To quantitatively demonstrate the consequence of the formal difference between different stress formulas, MD simulation results of the stress field near the dislocation core are provided.

## II. EXISTING MICROSCOPIC STRESS FORMULAS AND THEIR INCONSISTENCIES

The fundamental link between a local density (e.g., mass or energy density) in the physical space and a dynamic quantity in phase space is through the use of either the infinitely-peaked Dirac delta function[26-29] or a smeared out version of the Dirac delta function, the latter is also referred to as localization function or weighting function[30]. Densities localized at a point were called point functions by Irving and Kirkwood, such as mass density and momentum density, "since mass or momentum of any molecule may be considered as localized at that molecule"[26]. These point function definitions of local densities were ensemble-averaged (by repeating the observations many times) in the early formulations[26-28]. It was noted by Irving and Kirkwood that these point functions, "though averaged neither over space nor time, satisfy equations that are identical in form to the equations of hydrodynamics".[26]

The internal force density can be defined as a point function. For classical interacting particles system in which the net force acting on an atom can be expressed as the vector sum of each of the interaction forces, the ensemble-averaged internal force density $f_{\text{int}}(x)$ can be expressed in terms of the difference between two δ-functions:

$$f_{\text{int}}(x) = \langle \sum_{k,l} F_{kl} \delta(r_k - x) \rangle = \langle \frac{1}{2} \sum_{k,l} F_{kl} (\delta(r_k - x) - \delta(r_l - x)) \rangle, \quad (1)$$

where $r_k$ is the position vector of particle $k$, and $F_{kl}$ is the interaction force between particles $k$ and $l$. The potential part of stress follows from writing $f_{\text{int}}(x)$ as a divergence,

$$f_{\text{int}}(x) = \nabla \cdot \sigma_{\text{pot}}(x). \quad (2)$$

Equation (2) is the stress - force relationship that has been proven to be valid at the microscopic quantum mechanics level[3, 31-33], the classical mechanics level[34], and the macroscopic continuum mechanics level[35]. Interpretation of Eq. (1) is, therefore, a critical element in all the atomistic formulations of local stress.

In the work of Irving and Kirkwood and many later developments, the δ-function in Eq. (1) is the Dirac delta function[26-29, 31-33]. To derive formulas for local stress and heat flux, the difference between the two δ-functions in Eq. (1) was expanded in a power series[26, 33, 36, 37]:

$$\delta(r_k - x) - \delta(r_l - x) = \nabla_x \cdot \left\{ r_{kl} + \frac{r_{kl}}{2!}(r_{kl} \cdot \nabla) + ... \right\} \delta(r_k - x), \quad (3)$$

where $r_{kl} = r_k - r_l$. Keeping only the first term of this series, the approximated local stress can be written as a sum of a kinetic part and a potential part:

$$\tilde{\sigma}(x) = -\langle \sum_k m_k \tilde{v}_k \tilde{v}_k \delta(r_k - x) \rangle - \frac{1}{2} \langle \sum_{k,l} F_{kl} r_{kl} \delta(r_k - x) \rangle, \quad (4)$$

where $\tilde{v}_k = v_k - v$; $v_k$ and $v$ are the particle velocity and the velocity field, respectively.

In the formulation of Hardy[30], ensemble averaging was not used, and the Dirac δ in Eq. (1) was replaced with a localization function, denoted by Δ, that is defined with a finite size and interpreted as "defining the region averaged over". To identify the local stress, Hardy[30], Noll[38], and later many others[34, 39-42] used the integral representation for the difference between two δ- or Δ-functions:

$$\Delta(r_k - x) - \Delta(r_l - x) = -\nabla_x \cdot r_{kl} \int_0^1 \Delta(r_k \lambda + r_l(1-\lambda) - x) d\lambda. \quad (5)$$

This leads to a close-form formula for the local stress[30],

$$\sigma^{\text{Hardy}} = -\sum_k m_k \tilde{v}_k \tilde{v}_k \Delta(r_k - x) - \frac{1}{2} \int_0^1 \sum_{k,l} F_{kl} r_{kl} B(k,l,x) d\lambda, \quad (6)$$

where $B(k,l,x) = \int_0^1 \Delta(r_k \lambda + r_l(1-\lambda) - x) d\lambda$ in Eq. (6) was called *bond function* by Hardy[30]. Hardy's formulation leaves one with the arbitrariness of the localization function. Impulse function, Gaussian, etc., were assumed in various atomistic simulations[30, 43]. The bond function, however, was not obtained from the localization function; instead, an assumption has to be made independent of the form of the localization function. Although it yields a close-form local stress formula, the formulation does not completely resolve the difference between atomistic stress and Cauchy stress.

Averaging Hardy's local stress in Eq. (6) or the approximate stress in Eq. (4) over the volume of the entire system, or a volume whose radius exceeds the cut-off of the potential, one recovers the virial stress formula[6, 7]:

$$\bar{\sigma}^{\alpha\beta} = -\frac{1}{V}(\sum_k m_k \tilde{v}_k^\alpha \tilde{v}_k^\beta + \frac{1}{2} \sum_{k,l} F_{kl}^\alpha r_{kl}^\beta). \quad (7)$$

Because the virial stress is formally written as a sum over atoms, each individual term in the formula has been taken to describe the local stress at an atom and is usually referred to as the *atomic virial stress*[9]. Averaging the approximated atomistic stress in Eq. (4) over an atomic volume, the stress formula becomes identical to the atomic virial stress. This is the stress formula that has been exclusively used to calculate local stress in molecular dynamics (MD) simulators[44]. Nevertheless, except for a few special conditions, this atomistic stress has been shown to differ from Cauchy stress[8-11]. The distinction can be demonstrated using a simple linear chain model of stationary atoms used by Tsai[8] and also by Cheung and Yip[9], as shown in Fig.2. Assuming a cross sectional area $A$, a uniform interatomic spacing $b$, and second nearest neighbor interaction, the atomic virial stress at atom $k$, the hardy's stress, and the Cauchy stress due to the interaction forces on the planes to the left and the right of atom $k$ are, respectively,

$$\sigma_{\text{virial}}^{xx}(x = r_k) = -\frac{1}{2V} \sum_l F_{kl} r_{kl} = \frac{1}{2A}(2F_{k-2,k} + F_{k-1,k} + F_{k,k+1} + 2F_{k,k+2})$$
$$\sigma_{\text{Hardy}}^{xx}(x = r_k) = \frac{b}{V}(F_{k-2,k} + F_{k-1,k} + F_{k,k+1} + F_{k,k+2} + F_{k-1,k+1}), \quad (8)$$

$$\sigma_{\text{Cauchy}}^{xx}(x = r_k-) = \frac{1}{A} \sum F = \frac{1}{A}(F_{k-2,k} + F_{k-1,k} + F_{k-1,k+1})$$
$$\sigma_{\text{Cauchy}}^{xx}(x = r_k+) = \frac{1}{A} \sum F = \frac{1}{A}(F_{k-1,k+1} + F_{k,k+1} + F_{k,k+2}). \quad (9)$$

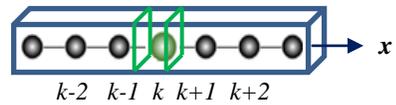

FIG. 2 A linear chain model for comparing atomistic and mechanical stress expressions.

This is a simple one-dimensional model and yet it clearly demonstrates that the differences between the results given by three local stress expressions are formal and irreconcilable, and can be significant when there is inhomogeneity resulted from, e.g., dislocations or cracks. This result is not unexpected since the virial stress was developed for the average stress of a total quantum or classical many-body system and hence it is not appropriate for atomic-level local stress. As for the Irving and Kirkwood's series representation, there is no proof that it converges at the atomic scale for general atomistic systems.

Note that equations (8) and (9) also show a fundamental formal difference between existing atomistic stress formulas and the concept of Cauchy stress: the former is defined as a volume-average, whereas the latter is simply area-averaged forces. The difference between various volume-average formulas, on the other hand, lies in the interpretation of Eq. (1).

## III. A NEW ATOMISTIC FORMULA FOR STRESS

To resolve both differences, we employ a different mathematic representation for the difference between two scalar-valued δ-functions. The *fundamental theorem for line integrals*, also known as the *gradient theorem*, states that "a line integral through the gradient of a scalar-valued function can be evaluated by evaluating the original scalar field at the endpoints of the line". It follows from the gradient theorem that

$$\delta(r_k - x) - \delta(r_l - x) = \int_{r_l}^{r_k} \nabla_\varphi \delta(\varphi - x) \cdot d\varphi = -\nabla_x \cdot \int_{L_{kl}} \delta(\varphi - x) d\varphi, \quad (10)$$

where $L_{kl}$ is a line from $r_l$ to $r_k$. Using Eqs. (1), (2), (10) the point function potential stress can be written as

$$\sigma_{\text{pot}}^{\text{point}}(x) = -\frac{1}{2} \int_{L_{kl}} \sum_{k,l} F_{kl} \delta(\varphi - x) d\varphi. \quad (11)$$

Equation (10) is mathematically valid for both the infinitely-peaked Dirac delta function and finite-sized localization functions. However, although particles are represented by points in a discrete atomistic system, a local density measures a physical quantity per unit volume or per unit area. For example, the particle density is defined as the number of particles per unit volume[33],

$$\rho_N(x) = \lim_{\Delta x \to 0} \frac{N(x, \Delta x)}{V(x, \Delta x)} = \sum_{k=1}^{n} \Delta(r_k - x). \quad (12)$$

It is seen from Eq. (12) that the $\Delta$-function is not infinitely peaked but has a maximum value of $1/V(x)$, where $V(x)$ is the volume of the domain at point $x$ that contains only one particle, i.e., $n = 1$. In general, for local densities that measure a physical quantity per unit volume, the Dirac δ must be averaged over a volume element. Denote the volume-averaged δ as $\bar{\delta}_V$; we have

$$\bar{\delta}_V(r_k - x) = \frac{1}{V(x)} \int_{V(x)} \delta(r_k - x') d^3 x' = \begin{cases} 1/V(x) \text{ if } r_k \in V(x) \\ 0, \text{ otherwise} \end{cases}. \quad (13)$$

Clearly, $\bar{\delta}_V(r_k - x)$ defined in Eq. (13) is consistent with the localization function defined by Hardy[30]: it peaks at $x = r_k$, satisfies $\int_V \bar{\delta}_V(r_k - x) dx = 1$, and is zero if particle $k$ is outside the domain surrounding the point $x$. However, it differs from Hardy's localization function by having a unique and definite value of $1/V(x)$. Volume elements with this value of volume can continuously fill the space the material system occupies, and consequently give rise to continuously varying local properties.

Similarly, for local properties that are by definition a measure of physical quantities per unit area, such as stress and heat flux vectors, we must average the Dirac δ over a surface element. Averaging Eq. (11) over a surface element $S_n(x)$ that centers at $x$ with area $A_n(x)$ and normal $n$, we obtain the stress vector $t(x, n)$ on $S_n(x)$:

$$\begin{aligned} t(x, n) &= -\frac{1}{A_n(x)} \int_{S_n(x)} n d^2 x' \cdot \frac{1}{2} \sum_{k,l} F_{kl} \int_{L_{kl}} d\varphi \delta(\varphi - x') \\ &= -\frac{1}{2} \sum_{k,l} F_{kl} \int_{L_{kl}} \bar{\delta}_A^n(\varphi - x) d\varphi \end{aligned}, \quad (14)$$

in which $\bar{\delta}_A^n(\varphi - x)$ is the area-averaged Dirac δ over $S_n(x)$.

The advantage of having Dirac delta in the formulation is that its concept is mathematically rigorous. For the line integral in Eq. (14) to be nonzero, one must have $\varphi - x' = 0$. That is, there must be a point $x'$ in $S_n(x)$ that lies on the line $L_{kl}$ (cf. Fig.3). Mathematically, this can be expressed as

$$\int_{L_{kl}} \bar{\delta}_A^n(\varphi - x) d\varphi = \frac{1}{A_n(x)} \int_{S_n(x)} n d^2 x' \cdot \int_{L_{kl}} d\varphi \delta(\varphi - x')$$
$$= \frac{1}{A_n(x)} \begin{cases} 1, \text{ if } x' \in S_n(x) \text{ and } x' \in [r_l \ r_k] \\ -1, \text{ if } x' \in S_n(x) \text{ and } x' \in [r_k \ r_l] \\ 0, \text{ otherwise} \end{cases}. \quad (15)$$

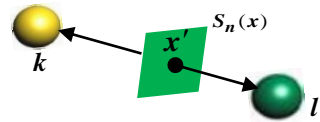

FIG. 3 A line segment connecting particles $k$ and $l$ intersects a surface element at $x'$, $x' \in S_n(x)$.

The line integral with the area-averaged δ in Eq. (14) now has a definite value and a clear physical meaning: for every particle-particle interaction, if the line segment $L_{kl}$ intersects the surface element $S_n(x)$, the contribution of the interaction force $F_{kl}$ to the stress vector on the surface element is then $F_{kl}/A_n(x)$, which is the interaction forces transmitted per unit area across the surface element $S_n(x)$.

Note that there is an infinite number of planes that pass through point $\boldsymbol{x}$. If we only consider three coordinate planes and denote $\boldsymbol{e}^\beta$ ($\beta = 1,2,3$) as the orthonormal basis at $\boldsymbol{x}$, we obtain the stress tensor in the indicial notation:

$$\sigma_{pot}^{\alpha\beta}(\boldsymbol{x}) = t^\alpha(\boldsymbol{x},\boldsymbol{e}^\beta) = -\frac{1}{2}\sum_{k,l}F_{kl}^\alpha \int_{L_{kl}} \bar{\delta}_A^\beta(\boldsymbol{\varphi}-\boldsymbol{x})d\boldsymbol{\varphi}. \quad (16)$$

Recall that the Cauchy stress tensor $\sigma^{\alpha\beta}$ is the α-th component of the stress vector acting on the β-th coordinate plane[12]. Clearly, the potential part of the atomistic stress tensor defined in Eq. (16) is unequivocally identical to the Cauchy stress.

The significance of the stress tensor defined in Eq. (16) is that by knowing the stress vectors on three mutually perpendicular planes, one can determine the stress vector on any other plane passing through the same point using *Cauchy tetrahedron argument*[45], i.e., $\boldsymbol{t}(\boldsymbol{x},\boldsymbol{n}) = \boldsymbol{n}\cdot\boldsymbol{\sigma}_{pot}(\boldsymbol{x})$. Note that the mechanical stress vector acting on the opposite sides of the same surface at a given point are equal in magnitude and opposite in sign, cf. FIG.4. This is the *Cauchy reciprocal theorem*, which is the counterpart of Newton's third law of motion that action is equal to reaction. Thus, it is essential for the potential (mechanical) stress to be defined as a planar average.

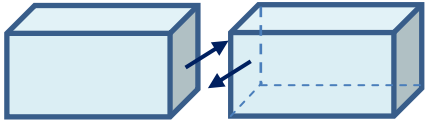

FIG.4 Stress vectors on the opposite sides of a surface.

## IV. A NUMERICAL EXAMPLE

To provide a quantitative understanding of the consequence, we compare in FIG.5 the stresses near a dislocation core calculated using the atomic virial stress formula with that using the new stress formula. The MD simulation is performed using LAMMPS[44]. The computer model is a two-dimensional Lennard-Jones single crystal that contains a stationary dislocation.

In FIG. 5 (a-c) we plot the potential stress distribution along the Y- and X- axes, and in Fig.5 (d) we present the virial stress per atom. It is seen that the stresses sharply increase as the measuring location approaches the dislocation core. However, the stress is not singular but has finite value, different from the prediction of the linear elasticity that ignores the structure of the dislocation core and the nonlocal atomic interaction. A dislocation produces a structural discontinuity that gives rise to a stress discontinuity. This is captured by the atomic virial stress formula. However, for a stationary dislocation, the atomic virial stress underestimates the stress by respectively 27%, 65%, and 38% for the stress components plotted in FIG.5.

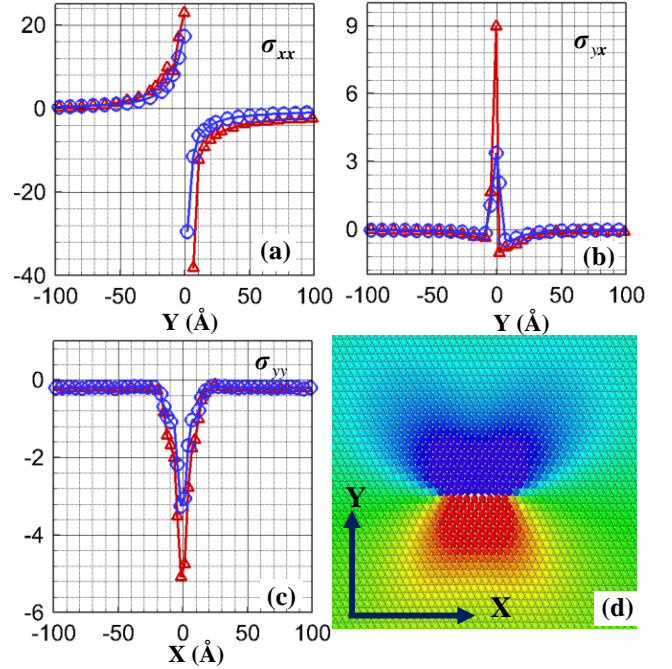

FIG. 5 Stresses near the dislocation core by atomic virial stress (circles, blue) using Eq. (4) and the Cauchy stress (triangles, red) using Eq. (17); the X-axis is in the direction of the Burgers vector and is above the dislocation core.

Since Hardy's stress is a volume average and depends on the volume to be averaged over as well as on the bond function to be assumed[23, 25], it is not calculated in this work. A detailed comparison between Hardy's stress and virial stress with different size of the analysis volume can be found in ref [25]. With an analysis volume larger than the volume of an atom, both the virial stress and the Hardy's stress predict a zero stress at the dislocation core, as shown in FIG. 1. These results demonstrate the consequence of the formal difference between the Cauchy stress and the existing microscopic stress formulas.

## V. DISCUSSIONS AND CONCLUSIONS

While Eq.(16) provides a new formula for the potential part of local stress, it can also be derived from Hardy's formula[30] by defining Hardy's localization function as an averaged Dirac delta function over an oriented surface:

$$\begin{aligned}\sigma_{pot}^{\alpha\beta}(\boldsymbol{x}) &= -\frac{1}{2}\int_0^1 \sum_{k,l} F_{kl}^\alpha r_{kl}\bar{\delta}_A^\beta\left(\boldsymbol{r}_k\lambda + \boldsymbol{r}_l(1-\lambda) - \boldsymbol{x}\right)d\lambda \\ &= -\frac{1}{2}\int_{\lambda=0}^{\lambda=1}\sum_{k,l}F_{kl}^\alpha \bar{\delta}_A^\beta\left(\boldsymbol{r}_{kl}\lambda+\boldsymbol{r}_l-\boldsymbol{x}\right)\frac{d}{d\lambda}\left(\boldsymbol{r}_{kl}\lambda+\boldsymbol{r}_l\right)d\lambda \\ &= -\frac{1}{2}\sum_{k,l}\int_{L_{kl}}F_{kl}^\alpha \bar{\delta}_A^\beta(\boldsymbol{\varphi}-\boldsymbol{x})d\boldsymbol{\varphi}\end{aligned} \quad (17)$$

Recall that in his Principles of Quantum Mechanics[46], Dirac listed properties of the delta function, one of which is $\delta(ax) = a^{-1}\delta(x)$. Clearly, Eq. (17) is consistent with this property. Thus, the stress formula of Hardy[30] can be consistent with the concept of Cauchy stress only if being interpreted through Eq. (17).

Irving and Kirkwood linked between the discrete description of molecular systems and the field equations of hydrodynamics by expressing local densities as point functions using Dirac δ-function[26]. Those point functions reflect the atomic granularity of matter, but must be averaged in space to describe local densities that are defined per unit volume or per unit area. Hardy used finite-size localization functions to define local densities, but did not distinguish between local densities defined per unit volume with those defined per unit area. This is the origin of the inconsistencies for volume-averaged stress formulas. Using the fundamental theorem for line integrals and area-averaged Dirac δ-function, we find a formal solution for the potential (or mechanical) stress that is fully consistent with the concepts of Cauchy stress, recover the Cauchy stress vector and stress tensor relationship, and satisfy the conservation equation of linear momentum.

A key difference between the new and existing atomistic formulas for the mechanical stress is that the former is a planar average, while the latter is a volume average. This difference is fundamental since the Cauchy stress is the actual force per unit area transmitted across a surface element in the deformed configurations. The planar average is a basic property of fluxes in general, and in particular is what enables one to quantify the stresses on specific planes such as cleavage planes and slip planes for understanding the physics of materials failure.

The obtained stress formulas follow necessarily from Eq. (1), i.e., the microscopic definition of internal force density, which is valid for classical interacting particle systems with non-polarizable (additive) force fields. Since no assumption is made on the type of the interaction forces, the new stress formula is valid for systems involving general many-body interactions. For multi-atom crystalline materials, the atomic volume element that contains only one atom should be replaced by one that contains more than one atom but one lattice point[39]. A two-level structural description is then more appropriate. Nonetheless, for both monatomic and multi-atomic crystals, the smallest volume of the volume element is that of the primitive unit cell.

**ACKNOWLEDGMENTS**

This material is based upon research supported by the U.S. DOE, Office of Basic Energy Sciences, Division of Materials Sciences and Engineering under Award # DE-SC0006539. We thank Ji Rigelesaiyin for the molecular dynamics simulation results. Discussions with Professors James Dufty, Simon Phillpot, Robert Hardy, and William Hoover on this work are gratefully acknowledged.

———————